\documentclass{elsart}
\usepackage{amssymb}
\usepackage{graphicx}

\usepackage{graphicx}

\begin{document}

\begin{frontmatter}

\title{{\em xPerm}: fast index canonicalization \\
       for tensor computer algebra}
\author[a]{Jos\'e M. Mart\'{\i}n-Garc\'{\i}a}%
\address[a]{Instituto de Estructura de la Materia, CSIC, \\
C/ Serrano 123, Madrid 28006, Spain}

\begin{abstract}
We present a very fast implementation of the Butler-Portugal algorithm
for index canonicalization with respect to permutation symmetries.
It is called {\em xPerm}, and has been written as a combination of a
{\em Mathematica} package and a C subroutine. The latter performs the
most demanding parts of the computations and can be linked from any other
program or computer algebra system.
We demonstrate with tests and timings the effectively polynomial
performance of the Butler-Portugal algorithm with respect to the number
of indices, though we also show a case in which it is exponential.
Our implementation handles generic tensorial expressions with several
dozen indices in hundredths of a second, or one hundred indices in a few
seconds, clearly outperforming all other current canonicalizers.
The code has been already under intensive testing for several years
and has been essential in recent investigations in large-scale
tensor computer algebra.
\end{abstract}


\end{frontmatter}

{\footnotesize
\section*{Program summary}
\textit{Title of program:} xPerm
\\
\textit{Catalogue identifier:}
\\
\textit{Program obtainable from:} (submitted to Computer Physics Communications) \\
\mbox{}\qquad \texttt{http://metric.iem.csic.es/Martin-Garcia/xAct/}
\\
\textit{Reference in CPC to previous version:} 
\\
\textit{Catalogue identifier of previous version:}
\\
\textit{Does the new version supersede the original program?:}
\\
\textit{Computers:} Any computer running C and/or Mathematica
\\
\textit{Operating systems under which the new version has been
tested:} Linux, Unix, Windows XP, MacOS
\\
\textit{Programming language:} C and Mathematica (version 5.0 or higher)
\\
\textit{Memory required to execute with typical data:} 20 Mbyte
\\
\textit{No. of bits in a word:} 64 or 32
\\
\textit{No. of processors used:} 1
\\
\textit{No. of bytes in distributed program, including test data,
etc.:} 1.5 Mbyte
\\
\textit{Distribution format:} Unencoded compressed tar file
\\
\textit{Nature of physical problem:} Canonicalization of indexed
expressions with respect to permutation symmetries.
\\
\textit{Method of solution:} The Butler-Portugal algorithm.
\\
\textit{Restrictions on the complexity of the problem:}
Multiterm symmetries are not considered.
\\
\textit{Typical running time:} A few seconds with generic expressions
of up to 100 indices.
}


\section{Introduction}
Tensor calculus is essential in those areas of Physics and Engineering
using curved manifolds and other geometrical structures to model
physical objects.
In particular, even the space-time itself can be the object of study,
as in General Relativity or other modern theories of gravity.
Computations in this context frequently become difficult, typically
due to two main sources of complexity: a high dimensionality $d$ or
working with a large number $n$ of indices.
This is because the size of coordinate expansions grows as $n^d$ and
the problem of symmetry handling grows as $n!$, respectively.
In those cases the use of Tensor Computer Algebra (TCA in the following)
can be of great help.
TCA is not only, however, a way of checking or avoiding our hand
computations; the use of efficient algorithms combined with moderate
computer resources (mainly RAM memory) can widen up enormously the
set of problems that we can attack and solve. For a general review on
TCA see Ref. \cite{MacCallum}.

This article focuses on the treatment of symmetries of tensorial
expressions under permutations of their indices.
Though the basic tensors used in our computations normally do not
have complicated symmetries, their arbitrary tensor products can be
very complicated.
Traditional methods, based on listing permutations, are exponential
in nature with respect to the number of indices, both in the time
and memory required to solve a problem. They become therefore rather
slow when working with expressions of several dozens of indices.
The Butler-Portugal \cite{Renato} algorithm, based on well-known
efficient methods of computational group theory, is effectively
polynomial and performs much better, as we will show with explicit
examples comparing with other systems.

Here we present for the first time a new implementation of the
Butler-Portugal algorithm.
It has been written both as a {\em Mathematica} package for generic
manipulation of large groups of permutations, and as a C subroutine
specialized in the index canonicalization problem. We shall refer
to these two parts of the code as {\em xPerm-Math} and {\em xPerm-C}
respectively.
This system is the computational core for a fully-fledged abstract
tensor package named {\em xTensor} by the same author, and both are
part of the integrated framework {\em xAct} \cite{xAct} for Tensor
Computer Algebra.
The properties and capabilities of {\em xTensor} will be reported
elsewhere.

Section \ref{algorithms} summarizes very briefly the theoretical
background behind the use of computational group theory in the problem
of index canonicalization. Sections \ref{xPerm-Math} and \ref{xPerm-C}
present the code, giving some examples of use. Section \ref{timings}
analyzes three nontrivial problems, showing the high efficiency and
speed of the system. Section \ref{conclusions} contains our conclusions.

\section{The algorithms}
\label{algorithms}


In this article a tensor, like $T^{cB1a}{}_{11c}$, is any expression
with a {\em head} symbol ($T$ in this case) and a list of {\em slots}
where we can place {\em indices}.
The expression can then be represented as {\tt T[c,B,1,a,-1,-1,-c]},
where we use a minus sign to denote covariant indices.
We shall not be concerned here with the types of indices that can be
used, but we shall assume the following: paired symbols (like {\tt c}
and {\tt -c} above) represent a contraction, in the usual Einstein
convention; integer indices do not follow the Einstein convention,
and can be repeated (like {\tt -1} above). The motivation for this is
using symbols to denote abstract indices and integers for components.

Any index configuration can be sorted into a canonical order following
several predefined priorities (for instance symbols before integers,
lowercase before uppercase, free before dummy, contravariant before
covariant, etc).
Following \cite{Renato}, we represent that index configuration as
the permutation which brings it to canonical order.
For example, if we assume that the canonical order for our
configuration $C=${\tt \{c,B,1,a,-1,-1,-c\}} is
$C_0\equiv${\tt \{a,c,-c,B,1,-1,-1\}} then $C$ can be converted into $C_0$
using the permutation $g=(1,2,4)(3,5,6,7)$, in disjoint cyclic notation;
that is, at the first slot of $C$ we find the second index of $C_0$,
at the second slot we find the fourth index, at the fourth slot we
find the first index, etc.
We follow the convention of applying permutations on the right:
$C_0= C \cdot g$. For example $(13)\cdot(132) = (12)$ and 
$(132)\cdot(13)= (23)$. In other words, $s^g = i$ means that the
image of slot $s$ under the permutation $g$ is the index $i$ in the
canonical list.
Experimentally we find that the choice of index priorities for the
canonical configuration deeply affects the overall efficiency of the
algorithm; having paired indices in consecutive slots of $C_0$
seems to be the best option.


A {\em permutation-symmetry} of a tensor is any rearrangement under
which the tensor remains invariant or changes sign. This forces us to
work with {\em signed permutations}, which belong to the extended group
$\{1, -1\}\times S_n$, rather than the symmetric group of $n$ indices
alone.
A signed permutation $(\sigma, \pi)$ with $\sigma = \pm 1$ and $\pi\in S_n$
acts on a tensor $T_{a_1\ldots a_n}$ producing
$\sigma\, T_{\pi(a_1 \ldots a_n)}$.
There are two types of permutation-symmetries:

\begin{itemize}
\item Those which involve a permutation of the slots independently
of which indices they have (let us call them {\em slot-symmetries}).
For the Riemann tensor examples are
\begin{displaymath}
R_{bacd} = - R_{abcd}, \qquad R_{cdab} = R_{abcd} .
\end{displaymath}
The slot-symmetries of a tensor form a permutation group, denoted
by $S$, not Abelian in general.
\item Those which involve a permutation among the indices
independently of which slots they are in (call them
{\em index-symmetries}). There are three possible sources:
\begin{itemize}
\item Exchange of dummies: $R^{ab}{}_{ab} = R^{ba}{}_{ba}$.
\item ``Metric swapping'': $R^{ab}{}_{ab} = R_a{}^{ba}{}_b$ (or
$\xi^A{}_A = -\xi_A{}^A$ for an antisymmetric metric, as in spinor
theory).
\item Repeated indices can always be exchanged, as the component
indices 1 in $R^{1a1b}$.
\end{itemize}
These three sources give rise to three respective groups $D_E$, $D_M$
and $D_R$, not Abelian in general, but which commute with each other.
The complete group of index-symmetries of a tensor
is $D=D_E\cdot D_M\cdot D_R$.
\end{itemize}

As shown in \cite{RenatoProducts}, every problem of index
canonicalization can be reduced to the problem of canonicalization
of the indices of single tensors: terms in a sum are treated
independently and factors in a tensor product can be sorted into
canonical order so that their indices can be considered to belong
to a single tensor. That tensor will inherit the slot-symmetries of the
factors in the product, but can have additional slot-symmetries coming
from exchange of several instances of the same tensor. (At this
point it is relevant whether the tensor product is commutative
or not.) In the following we shall assume that we work with a single
tensor $T^{\ldots}{}_{\ldots}$ of arbitrary slot-symmetry group $S$ and
index-symmetry group $D$.


A given configuration of indices $g$ is equivalent to any other
configuration $s\cdot g\cdot d$ with $s\in S$ and $d\in D$. The set
$S\cdot g\cdot D$ of permutations is called a {\em double coset}, not
a group in general, and the problem of canonicalization of a
configuration $g$ in the presence of symmetries $S$ and $D$ can be
reduced to defining and finding a canonical representative for the
corresponding double coset\footnote{As far as I know, this was first
elaborated in Ref. \cite{russians} and implemented in {\em Reduce}.}.
Such an algorithm was given by Butler \cite{Butler_doublecosets}
for generic groups $S$ and $D$, making intensive use of the efficient
group representation provided by a {\em strong generating set}
\cite{Butler_book}:
Given a group $G$ of permutations acting on a set $\Omega$ of points,
a subset $B=\{b_1, ..., b_k\}$ of points $b_i\in \Omega$ is said to be
a {\em base} of $G$ iff none of the permutations of $G$ fixes
all points in B. Then we say that a generating set $\Delta$ of $G$
is {\em strong} with respect to $B$ iff there is a generating set
in $\Delta$ for each of the stabilizer subgroups
$G_i = \{g \in G ,,\  b_j{}^g = b_j \ {\rm for}\ j=1...i\}$. This
induces a hierarchy $G=G_0 \supseteq G_1 \supseteq ... \supseteq G_k
= \{{\rm id}\}$ which efficiently represents the group $G$. The key
idea is that problems concerning permutations in $G$ can be translated
into equivalent problems for the cosets of the $G_i$ in $G_{i-1}$.
A strong generating set and its base can be obtained from a
normal generating set using the Schreier-Sims algorithm
\cite{Butler_book}.

Butler's algorithm \cite{Butler_doublecosets}
for a permutation $g\in G$ starts from respective
strong generating sets $(B_S,\Delta_S)$ and $(B_D,\Delta_D)$ of two
subgroups $S$ and $D$ of $G$. The base $B_S$ is extended to be a base
$B$ of $G$ and then sequentially the points of base $B_D$ are changed
in a way determined by $g$, so that in the end $B_D = B^p$, uniquely
identifying the canonical permutation $p$ of the double coset
$S\cdot g\cdot D$.
Butler's algorithm has been adapted by Portugal and
collaborators \cite{Renato} to the case of index canonicalization for
a generic group $S$ and a group $D$ of the form $D=D_E\cdot D_M$, which
simplifies the process because the properties of the strong generating
set for such a $D$ group are known in advance. In tensorial terms,
the base $B_S$ defines priorities among the slots, and then the
algorithm sequentially decides which indices are placed at those
slots. Zero is returned as soon as $-g$ is shown to belong to the
double coset $S\cdot g\cdot D$. From now on we shall
refer to this algorithm as the Butler-Portugal algorithm. For full
details and examples see Ref. \cite{Renato}, and the {\em Canon}
\cite{Canon} implementation in {\em Maple} by the same authors.

Butler's algorithm uses internally the intersection algorithm and hence
it has global exponential efficiency in the number of indices at hand.
In practical generic applications, however, it is effectively
polynomial because only in a very small subset of cases the exponential
character is apparent. This will become clear in the examples shown in
Sect. \ref{timings}.


In this article we perform two straightforward extensions
to include the possibility of working with indices of different types
(usually indices on different vector spaces) and also with component
indices, which introduce the possibility of being repeated.
Note that both issues affect the $D$ group, but not the $S$ group.
The key observation is that the $D_E$ and $D_M$ groups act only on
abstract indices, while the $D_R$ group acts only on component indices,
and so their actions are disjoint. The construction of the strong
generating set for $D$ from those of the three groups is hence trivial.
The same observation applies for the groups of the indices of different
types which, by definition, also have disjoint action.
Minor modifications of the Butler-Portugal algorithm are required to
handle the new $D_R$ groups.

\section{A {\em Mathematica} implementation}
\label{xPerm-Math}

The {\em xPerm} package contains a number of tools for efficient
manipulation of large groups of permutations. It has been fine tuned
to the problem of index canonicalization, but can be used in a more
general context. It is, however, rather limited in comparison with
general-purpose environments in Computational Group Theory like
MAGMA \cite{MAGMA} or GAP \cite{GAP}.

The package is loaded using standard {\em Mathematica} notation:

\qquad {\em In[1] :=\ } {\tt <<xAct`xPerm` } \\
\mbox{}\qquad\qquad\qquad - - - - - - - - - - - - - - - - - - - - - - - - - - - - - - - - - - - - - - - - - - \\
\mbox{}\qquad\qquad\qquad Package xAct`xPerm`\ \  version 1.0.0,
\{2008, 3, 5\} \\
\mbox{}\qquad\qquad\qquad Copyright (C) 2003--2008 Jose M. Martin-Garcia, under GPL \\
\mbox{}\qquad\qquad\qquad - - - - - - - - - - - - - - - - - - - - - - - - - - - - - - - - - - - - - - - - - -

There are three sets of commands and algorithms in {\em xPerm}.
First we have basic tools to manipulate permutations. To avoid problems
with the choice of notations, and to compare their different
efficiencies, four different notations have been implemented.
The permutation taking the numbers \{1,2,3,4,5,6\} to \{3,2,4,1,6,5\}
can be represented using either of these:
\begin{itemize}
\item[] {\tt Perm[\{3, 2, 4, 1, 6, 5\}]}, rearrangement of numbered objects.
\item[] {\tt Images[\{4, 2, 1, 3, 6, 5\}]}, list of images.
\item[] {\tt Cycles[\{1, 4, 3\}, \{5, 6\}]}, in disjoint cyclic notation.
\item[] {\tt Rules[1->4, 4->3, 3->1, 5->6, 6->5]}, convenient in
{\em Mathematica}.
\end{itemize}
The function {\tt TranslatePerm} changes among different notations.
Other simple functions include {\tt PermDeg}, {\tt InversePerm},
{\tt PermSort}, etc., with obvious meanings. Products of permutations
are taken with {\tt Permute}. Given a generating set of permutations,
the associated group can be constructed using the {\tt Dimino}
algorithm.

A second type of algorithms are those in charge of constructing and
manipulating strong generating sets. We have encoded a larger number
of the algorithms given in Ref. \cite{Butler_book}, which include a
combination of tools to analyze how points move under the permutations
(orbits and Schreier vectors) and how they do not move under them
(stabilizers).
Relevant commands are {\tt Orbit}, {\tt SchreierOrbit},
{\tt TraceSchreier}, {\tt Stabilizer}, and others.
The main algorithm is {\tt SchreierSims}, which constructs a strong
generating set of a group $G$ from any generating set of $G$.
For example, the symmetry group $S$ of the Riemann tensor can be
described by the generating set

\qquad {\em In[2] :=\ }
{\tt GS = GenSet[ -Cycles[\{1,2\}], Cycles[\{1,3\},\{2,4\}] ];}

The list \{1,3\} is a base for $S$, but {\tt GS} is not strong with
respect to it (there is no permutation of the stabilizer $S_1$ in
{\tt GS}). We construct a strong generating set using

\qquad {\em In[3] :=\ } {\tt SGS = SchreierSims[\{\}, GS]} \\
\mbox{}\qquad {\em Out[3] =\ } 
StrongGenSet[\,\{1,\,3\},  \\
\mbox{}\qquad\qquad\qquad\quad  GenSet[$-$Cycles[\{1,\,2\}],\,Cycles[\{1,\,3\},\,\{2,\,4\}],\,$-$Cycles[\{3,\,4\}]]\,]

which has added the missing permutation.
The group is then decomposed into a hierarchy of subgroups which can
be used to get any piece of information using the principles of
inductive foundation and backtrack search (the {\tt Search} command).
In particular we can test membership in the group with {\tt PermMemberQ}
and compute its order with {\tt OrderOfGroup}:

\qquad {\em In[4] :=\ } {\tt OrderOfGroup[ SGS ] } \\
\mbox{}\qquad {\em Out[4] = \ } 8

The documentation of {\em xPerm} shows all examples in
\cite{Butler_book} concerning these algorithms, correcting the results
of some of them.

The final third set of implemented algorithms are those needed for index
canonicalization, following \cite{Renato} with the simple extensions
mentioned before. There is {\tt RightCosetRepresentative}, which gives
the canonical representative of the right coset $S\cdot g$ for any
permutation $g$ and any group $S$. This is used to canonicalize free
indices first. Then there is {\tt DoubleCosetRepresentative}, which does
the same for the double coset $S\cdot g\cdot D$, canonicalizing the
dummy and component indices, taking into account the slot-symmetries
which do not move the slots with free indices. This two-step process
is more efficient than using only the second one only.
Finally the function {\tt CanonicalPerm} is in charge of the
combination of both algorithms and is the only function to be called
when doing index canonicalization. Let us see an example:
%
%

Consider the expression $R_b{}^{1d1}\,R_c{}^{bac}$.
A product of two Riemann tensors has the slot-symmetries described by

\qquad {\em In[5] :=\ }
{\tt SGS = StrongGenSet[ \{1,3,5,7\}, GenSet[ \\
\mbox{}\hspace{1cm}-Cycles[\{1,2\}],\,-Cycles[\{3,4\}],\,-Cycles[\{5,6\}],\,-Cycles[\{7,8\}],\\
\mbox{}\hspace{1cm}Cycles[\{1,3\},\{2,4\}],\,Cycles[\{5,7\},\{6,8\}],\\
\mbox{}\hspace{1cm}Cycles[\{1,5\},\{2,6\},\{3,7\},\{4,8\}] ] ] };

The first line of permutations encodes the four pair antisymmetries; the
second line describes symmetry under exchange of pairs in each Riemann;
the third line exchanges both Riemanns. The indices
{\tt \{a,d,b,-b,c,-c,1,1\}}, already in canonical order have
index-symmetries described with the notation:

\qquad {\em In[6] :=\ }
{\tt Dsets=\{DummySet[M,\{\{3,4\},\{5,6\}\},1],RepeatedSet[\{7,8\}]\};}

where the {\tt DummySet} expression contains all dummies of space {\tt M},
which has a symmetric metric (switch 1). There could be several
of those expressions, one for each vector space, and several
{\tt RepeatedSet} expressions, one per repeated index.
Other indices in the canonical list are free:

\qquad {\em In[7] :=\ }
{\tt frees = \{1,2\};}

Our expression is represented by the permutation

\qquad {\em In[8] :=\ }
{\tt perm = Cycles[\{1,4,8,5,6,3,2,7\}];}

Hence the expression can be canonicalized using

\qquad {\em In[9] :=\ }
{\tt CanonicalPerm[\,perm,\,8,SGS,\,frees,\,Dsets\,]} \\
\mbox{}\qquad {\em Out[9] =\ }
Cycles[\{2,3,4,5\},\{6,7\}]

which corresponds to the expression $R^{ab}{}_b{}^c\,R^{d1}{}_c{}^1$.

Canonicalization can be performed using pure {\em Mathematica} code
({\em xPerm-Math}) or linking to an external compiled executable
({\em xPerm-C}) through the {\em MathLink} protocol, which is faster.
This is controlled with the options\, {\tt MathLink -> False} or
{\tt MathLink -> True} of {\tt CanonicalPerm}, respectively.

\section{The C code for the canonicalizer}
\label{xPerm-C}

All algorithms required for index canonicalization have been recoded
in 2400 lines of C code to increase the speed the system. We use some
features of the C99 standard and hence the code must be
compiled with some modern C compiler like the GNU {\tt gcc} compiler
for Linux, Unix or Mac, or its {\tt cygwin} port to Windows.
For those systems with no such compilers the pure {\em Mathematica}
code presented in the previous section is always available, with
identical output, but slower.

Permutations are now represented always as lists of images (this is the
{\tt Images} notation in the {\em Mathematica} code of {\em xPerm}).
The sign of a permutation is encoded in an additional pair of points
at the end of the permutation. For instance {\tt Images[\{2,3,4,1\}]}
will be encoded as the list {\tt \{2,3,4,1,5,6\}} and 
{\tt -Images[\{2,3,4,1\}]} as {\tt \{2,3,4,1,6,5\}}.
The product of two permutations {\tt p1} and {\tt p2}, both of degree
{\tt n}, is efficiently performed with the function
\begin{verbatim}
    void product(int *p1, int *p2, int *p, int n) {
            while(n--) *(p++) = *(p2-1+*(p1++)) ;
    }
\end{verbatim}
which stores the result in {\tt p}. Similarly, the inverse {\tt ip} of a
permutation {\tt p} of degree {\tt n} is computed using
\begin{verbatim}
    void inverse(int *p, int *ip, int n) {
            while(n--) *(ip-1+*(p+n)) = n+1 ;
    }
\end{verbatim}
A generating set of $m$ permutations of degree $n$ is stored as a list
of $m\times n$ integers. Functions are provided to compute images of
points under permutations, stable points, stabilizers, orbits, Schreier
vectors and other basic constructions in theory of finite groups.

The Schreier-Sims algorithm has also been encoded in C, for
completeness, though in typical applications in tensor algebra we do
not expect to find the problem of computing a complex strong generating
set from a given set of permutations. That could happen, for example,
if we wanted to canonicalize a tensor with 48 indices and Rubik's group
as symmetry (a group of $4\cdot 10^{19}$ permutations).
Providing the obvious set of six generators it takes 5 seconds in
{\em xPerm} to compute a strong generating set for that group.
As a second nontrivial example, a strong generating set for the
smallest of the three Conway simple sporadic groups, represented
with permutations acting on 276 points, is constructed in 2.5 seconds.
(This is a group of order $5\cdot 10^{11}$.)

Finally there is the canonicalizer function, called
{\tt canonical\_perm\_ext}. Prototypes and further explanations on the
structure of this and other functions are given in Appendix 
\ref{prototypes}, to help linking from other codes.
In {\em Mathematica} the link can be performed via the {\em MathLink}
protocol, and a template ({\tt .tm}) file is provided with the
{\em xPerm} package for this purpose.

\section{Examples and timings}
\label{timings}

\begin{figure}[ht!]
\begin{center}
\includegraphics[width=14cm]{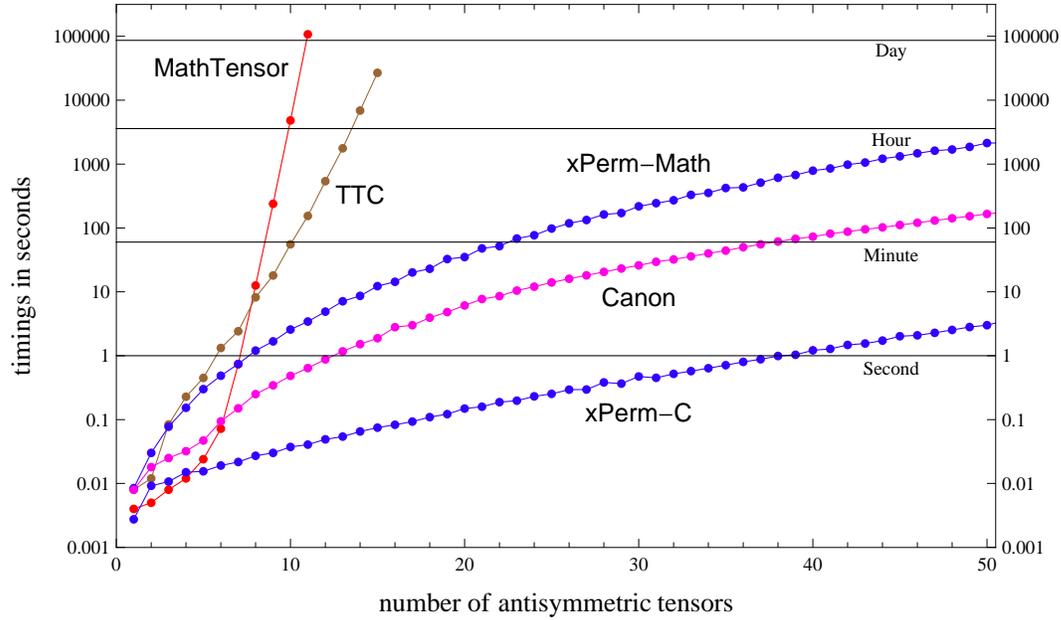}
\caption{\label{comparison}
Timings, in a logarithmic axis, for canonicalization of expression
(\ref{anti}) with various tensor canonicalizers: {\em MathTensor}
\cite{MathTensor}, {\em TTC} (Tools of Tensor Calculus) \cite{TTC},
{\em Canon} \cite{Canon} and both the pure {\em Mathematica} and
the C-based codes reported in this article. The exponential character
of the traditional methods vs. the effectively polynomial character
of the Butler-Portugal algorithm is apparent.
}
\end{center}
\end{figure}

{\bf Example 1}: Given a general antisymmetric tensor $F_{ab}= - F_{ba}$
it is simple to see that
\begin{eqnarray} \label{anti}
F^{a_1}{}_{a_2}\, F^{a_2}{}_{a_3}\,\cdots\,F^{a_n}{}_{a_1} 
= \left\{ \begin{array}{l} =0 \qquad {\rm if\ n\ odd} \\
                       \not=0 \qquad {\rm if\ n\ even} \end{array} \right.
\end{eqnarray}
The slot-symmetry group $S$ in this problem is generated by the
inherited permutations $-(1,2)$, $-(3,4)$, ..., $-(n-1, n)$ and the
tensor-exchange permutations $(1,3)(2,4)$, $(3,5)(4,6)$, ...,
$(n-3,n-1)(n-2,n)$, in disjoint cyclic notation.
This is a strong generating set with respect to the base points
$\{1,3,\ldots,n-1$\}.
By accident, the groups $S$ and $D$ coincide, though this does not
introduce any simplification in the problem.
These are large groups, with orders $5\cdot 10^{32}$ for $n=25$
and $3\cdot 10^{79}$ for $n=50$.
Figure \ref{comparison} shows a comparison of timings of
canonicalization for different systems with respect to the number $n$
of antisymmetric tensors. All systems correctly find that
expression (\ref{anti}) is zero for odd $n$, and it is clear that
{\em xPerm} and {\em Canon}, which use the Butler-Portugal algorithm,
are far more efficient than those systems using more traditional methods.
We see that {\em xPerm-C} can handle 100 indices within a few
seconds. Note also that more sophisticated methods carry an
overhead which makes them slower for very low $n$.

\begin{figure}[ht!]
\includegraphics[width=14cm]{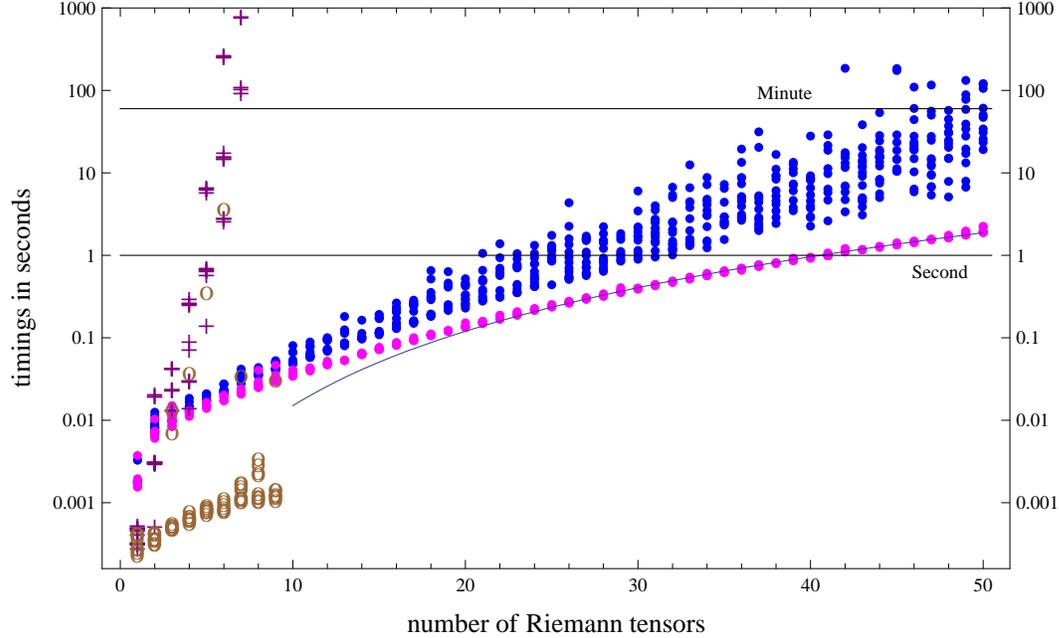}
\caption{\label{Riemann_comparison}
Timings, in a logarithmic axis, for canonicalization of scalar products
of $n$ Riemann tensors. There are 20 points per value of $n$ for {\em
MathTensor} \cite{MathTensor}, separated in those cases giving zero
(brown open circles) and those given nonzero result (purple crosses).
The exponential blow up is clear and it is not possible to go beyond
$n=9$ with 1Gbtyte RAM. There are also 20 points per value of $n$ for
the C canonicalizer of {\em xPerm} and we show results up to $n=50$
(that is, 200 indices), though it is possible to go much beyond with
that RAM memory.
Dark blue circles represent cases with nonzero results and
magenta filled circles are cases with zero result (for reference we
display the curve $t(n) = 15\, n^3\, \mu s$).
}
\end{figure}

{\bf Example 2:} We canonicalize random Riemann monomials with all
indices contracted.
Figure \ref{Riemann_comparison} shows the results for products of up
to 9 Riemanns in {\em MathTensor} \cite{MathTensor} and products of
up to 50 Riemanns with our C canonicalizer.
The global structure coincides with that of Figure \ref{comparison},
and in particular we also see that 100 indices can be also manipulated
in just a few seconds.
It is interesting to comment on the different behaviour of the cases
giving zero and those giving nonzero results.
{\em MathTensor} has a special rule transforming contractions of
type $R^a{}_{abc}$ into 0, and those trivial zeros are detected in
milliseconds.
However there are other nontrivial zeros, like $R^{ab}R_{abcd}$,
which take even more than a second. The systematic and general
algorithm of {\em xPerm} does not distinguish between trivial and
nontrivial zeros and detects both in the same way, with an efficiency
which seems to grow as $n^3$. The non-zero results show a large
disperssion in timings, but the average seems to grow like $n^5$,
confirming the result of a similar experiment in \cite{Renato}.

The canonicalization of Riemann monomials is essential in our recent
construction of the {\em Invar} package for fast manipulation of
Riemann invariants, both algebraic invariants \cite{Invar} and
differential invariants \cite{DInvar}.

\begin{figure}[ht!]
\begin{center}
\includegraphics[width=13cm]{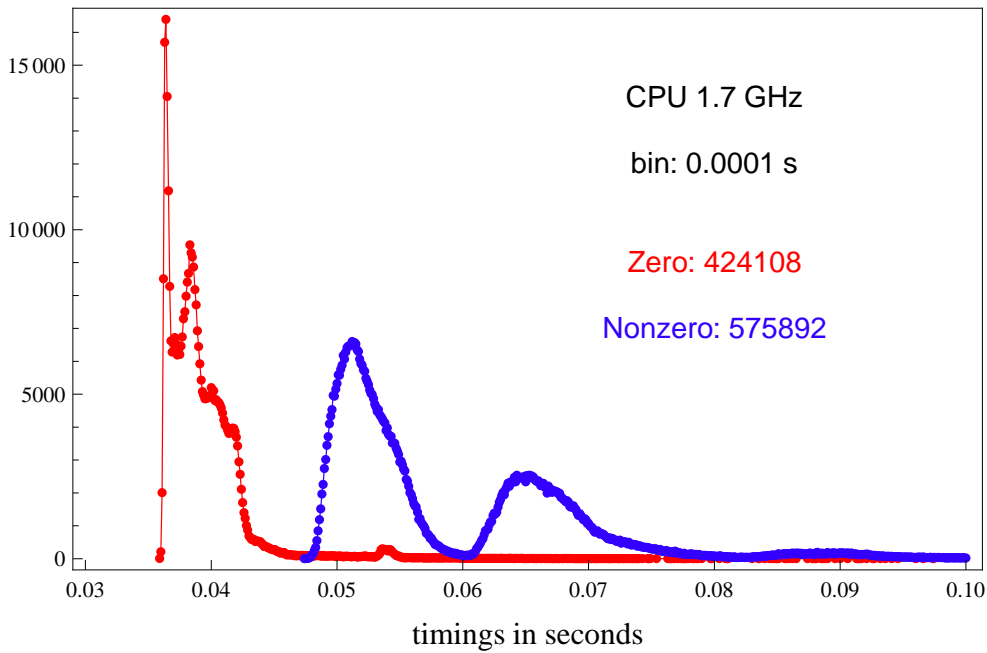}
\caption{\label{Riemann10}
Histogram of canonicalization timings of one million algebraic Riemann
invariants of degree 10 with our C canonicalizer.
Zeros (red) are found faster than nonzero (blue) results.
The former are all contained in the figure, but there are 4097 nonzero
cases taking more than 0.1 s, and actually six of those take 
between 1 and 6 seconds.
Note the origin of the timings axis at 0.03 s. This is roughly the
time that the driving tensor system {\em xTensor} \cite{xAct} takes to
compute the $S$ and $D$ groups from the tensor expression.
}
\end{center}
\end{figure}

{\bf Example 3:} Let us now concentrate on Riemann invariants of degree
$n=10$, so that we work with 40 indices.
Fig. \ref{Riemann10} shows an histogram of canonicalization timings
of one million random invariants, separating those cases giving zero
from those giving nonzero results. It is clear that zeros are faster
than nonzero results, because the Butler-Portugal algorithm gives 0 as
soon as $-g$ is found in the $S\cdot g\cdot D$ double coset. We also
see several peaks, corresponding to cases of different complexity.
All zero cases are found in less than 0.1s, while several cases
take more than a second to canonicalize.

We believe that the hardest Riemann invariants to canonicalize are
those of the form
\begin{equation}
R^{a_1b_1}{}_{a_2b_2}\, R^{a_2 b_2}{}_{a_3b_3}\,\ldots\,
R^{a_nb_n}{}_{a_1b_1}
\end{equation}
which require listing $2n\times 2^n$ permutations internally in 
the Butler-Portugal algorithm (in particular in the ALPHA table,
see \cite{Renato}), which clearly renders the whole process exponential
in $n$. Compare this with a mere $2n$ for expressions (\ref{anti}).
Once the hard cases have been identified it is always possible to
prepare the system in advance to detect them and switch to adapted
algorithms. For example,
in this case the antisymmetric pairs $ab$ in dimension $d$ can be
replaced by a single index $A$ in dimension $d(d-1)/2$ converting
the problem into $R^{A_1}{}_{A_2}\, R^{A_2}{}_{A_3}\, \ldots
R^{A_n}{}_{A_1}$, with $R_{AB}$ symmetric. This new problem is now
nearly identical in nature to our Example 1, and hence as efficient.

\section{Conclusions}
\label{conclusions}

We have presented {\em xPerm}, a free-software package for efficient
manipulation of large permutation groups in {\em Mathematica}.
Its main objective is index canonicalization under permutation symmetries
of tensorial expressions. For this, {\em xPerm} implements the
Butler-Portugal algorithm, which is effectively polynomial in the number
of indices, in contrast to the clear exponential behaviour of more
traditional algorithms.  As far as the author knows, {\em xPerm}
is currently the fastest index canonicalizer, and this has been
demonstrated comparing timings in some nontrivial example computations
with other tensor computer algebra systems.

With a tool like this at hand, we can enormously extend the type of
problems in differential geometry that we can attack and solve. There
are already several examples of projects in General Relativity which
have been possible only through the use of {\em xPerm}. To name a few:
the analysis of the hyperbolicity properties of large families of
formulations of the Einstein equations \cite{hyp} used in Numerical
Relativity; the construction of the dynamical evolution equations for
the Bel tensor \cite{superenergy}; the computation of 
the polarization of gravitational radiation at the third post-Newtonian
order \cite{3PN}; or the construction of a large database of equations
among the algebraic and differential invariants of the Riemann tensor
\cite{Invar,DInvar}.

Authors of other \underline{free-software} packages are encouraged to
link to the C code presented here to gain efficiency in the process
of index canonicalization, with the only proviso that they must require
their users to cite this publication any time the combination of
packages is used. The standalone system {\em cadabra} \cite{cadabra}
already has this code at its computational core.

{\em xPerm} can be downloaded from the webpage

$\qquad\qquad${\tt http://metric.iem.csic.es/Martin-Garcia/xAct/}

\section*{Acknowledgements}
I thank Paolo Matteucci for suggesting the use of the Butler-Portugal
algorithm for index canonicalization.
I also thank Alfonso Garc\'\i a-Parrado, Carsten Gundlach,
Kasper Peeters, Renato Portugal and David Yllanes for helpful
discussions and for their help testing the system.
Support from the Spanish MEC under the research project
FIS2005-05736-C03-02 is acknowledged.
Part of the computations have been performed at the {\em Centro de
Supercomputaci\'on de Galicia} (CESGA).

\appendix

\section{Prototypes of C functions}
\label{prototypes}

1) The Schreier-Sims algorithm: Given a generating set $\Delta$ of $m$
elements for the group $G\subseteq S_n$, return a strong generating
set $\Delta'$ with respect to a base $B$ for the group $G$:

\begin{verbatim}
void schreier_sims(int *base, int bl, int *GS, int m, int n,
        int *newbase, int *nbl, int **newGS, int *nm, int *num);
\end{verbatim}

with input

\mbox{}\qquad {\tt base}: list of the first points of the base, \\
\mbox{}\qquad {\tt bl}: length of the list {\tt base}, \\
\mbox{}\qquad {\tt GS}: list containing the original generating set, \\
\mbox{}\qquad {\tt m}: number of permutations in {\tt GS}, \\
\mbox{}\qquad {\tt n}: degree of the permutations,

and output 

\mbox{}\qquad {\tt newbase}: final list of points in the base, \\
\mbox{}\qquad {\tt nbl}: number of points in the base, \\
\mbox{}\qquad {\tt newGS}: strong generating set, \\
\mbox{}\qquad {\tt nm}: number of permutations in {\tt newGS}, \\
\mbox{}\qquad {\tt num}: number of generators checked in the process.

2) Order of a group: Given a strong generating set $\Delta$ with respect to
a base $B$, return the order of the group $G$ generated:

\begin{verbatim}
long long int order_of_group(int *base, int bl, int *GS, int m, int n);
\end{verbatim}

with identical input to the previous function and obvious output.

3) Membership in a group: Given a strong generating set $\Delta$ of $G$ with
respect to a base $B$ find whether a given permutation $g$ belongs to $G$.

\begin{verbatim}
int perm_member(int *p, int *base, int bl, int *GS, int m, int n);
\end{verbatim}

with identical input plus

\mbox{}\qquad {\tt p}: permutation to be tested.

and output 1 if the permutation {\tt p} belongs to the group $G$ or
0 otherwise.

4) Double coset canonicalization: Given a permutation $g$ we return its
canonical representative taking into account the information of the
slot-symmetry group $S$ and the free, dummy and repeated indices.

\begin{verbatim}
void canonical_perm_ext(int *perm, int n,
        int SGSQ, int *base, int bl, int *GS, int m,
        int *freeps, int fl,
        int *vds, int vdsl, int *dummies, int dl, int *mQ,
        int *vrs, int vrsl, int *repes, int rl,
        int *cperm);
\end{verbatim}

Input:

\mbox{}\qquad {\tt perm}: permutation to be canonicalized \\
\mbox{}\qquad {\tt n}: degree of all permutations in the computation \\
\mbox{}\qquad {\tt SGSQ}: 1 if {\tt GS} is strong and 0 otherwise \\
\mbox{}\qquad {\tt base}: base of the strong generating set of $S$ \\
\mbox{}\qquad {\tt bl}: length of the list {\tt base} \\
\mbox{}\qquad {\tt GS}: (strong) generating set of $S$ \\
\mbox{}\qquad {\tt m}: number of permutations in {\tt GS} \\
\mbox{}\qquad {\tt frees}: free indices in input \\
\mbox{}\qquad {\tt fl}: length of the list {\tt frees} \\
\mbox{}\qquad {\tt vds}: list of lengths of dummy-sets \\
\mbox{}\qquad {\tt vdsl}: length of list {\tt vds} \\
\mbox{}\qquad {\tt dummies}: list with pairs of dummies \\
\mbox{}\qquad {\tt dl}: length of list {\tt dummies} \\
\mbox{}\qquad {\tt mQ}: list of symmetries of metric (-1, 0, 1) \\
\mbox{}\qquad {\tt vrs}: list of lengths of repeated-sets \\
\mbox{}\qquad {\tt vrsl}: length of list {\tt vrs} \\
\mbox{}\qquad {\tt repes}: list with repeated indices \\
\mbox{}\qquad {\tt rl}: length of list {\tt repes}

Output:

\mbox{}\qquad {\tt cperm}: canonical permutation

For the example in Sect. \ref{xPerm-Math} the input data would as
follows. (Note the permutations degree $n=10$: the
two last points encode the permutation sign.)

\mbox{}\qquad {\tt perm = \{4,7,2,8,6,3,1,5,9,10\} \qquad \tt n = 10} \\
\mbox{}\qquad {\tt SGSQ = 1} \\
\mbox{}\qquad {\tt base = \{1,3,5,7\} \qquad bl = 4} \\
\mbox{}\qquad {\tt GS = \{2,1,3,4,5,6,7,8,10,9,\ \ 1,2,4,3,5,6,7,8,10,9,\\
\mbox{}\hspace{19mm} 1,2,3,4,6,5,7,8,10,9,\ \ 1,2,3,4,5,6,8,7,10,9,\\
\mbox{}\hspace{19mm} 3,4,1,2,5,6,7,8,9,10,\ \ 1,2,3,4,7,8,5,6,9,10,\\
\mbox{}\hspace{19mm} 5,6,7,8,1,2,3,4,9,10\} \qquad m = 7} \\
\mbox{}\qquad {\tt frees = \{1,2\} \qquad fl = 2} \\
\mbox{}\qquad {\tt vds = \{4\} \qquad mQ = \{1\} \qquad vdsl = 1} \\
\mbox{}\qquad {\tt dummies = \{3,4,5,6\} \qquad dl = 4} \\
\mbox{}\qquad {\tt vrs = \{2\} \qquad vrsl = 1} \\
\mbox{}\qquad {\tt repes = \{7,8\} \qquad rl = 2}

The output would be

\mbox{}\qquad {\tt cperm = \{1,3,4,5,2,7,6,8,9,10\} }

\end{document}